\documentclass[twocolumn]{aastex7}
\begin{document}

\title{The Cygnus X-1 Puzzle: Implications of X-ray Polarization Measurements in the Soft and Hard States on the Properties of the Accretion Flow and the Emission Mechanisms}

\author[orcid=0000-0002-1084-6507,gname='Henric',sname='Krawczynski']{Henric Krawczynski}
\affiliation{Washington University in St. Louis, the McDonnell Center for the Space Sciences and the Center for Quantum Leaps, St. Louis, MO 63130}
\email[show]{krawcz@wustl.edu}  
\author[orcid=0000-0002-9705-7948,gname='Kun',sname='Hu']{Kun Hu}
\affiliation{Washington University in St. Louis and the McDonnell Center for the Space Sciences, St. Louis, MO 63130}
\email[show]{hkun@wustl.edu}  

\begin{abstract}
In this paper, we summarize key observational constraints of the accretion flow on the black hole X-ray binary Cygnus X-1 (Cyg X-1). The discussion highlights the flows of energy close to the black hole and the importance of the distance range from which the radiating zone draws its energy. For the hard state, we examine compact and extended corona models. 
We find that compact corona models are energetically favored, but extended models cannot 
be fully excluded. We discuss the high linear polarization 
of the Cyg X-1 X-rays in the soft and hard states, parallel to the direction of the radio jet. 
We propose the presence of a pair layer enveloping the accretion disk moving at approximately half the speed of light 
away from the disk for both the soft and the hard state.
In the soft state, the pairs cool to the Compton 
temperature of the disk emission. In the hard state,
the pairs acquire thermal and bulk motion allowing 
them to Comptonize the emission to produce the observed 
power law emission. 
In both emission states, the bulk motion away from the disk leads to a net polarization parallel to the radio jet.
We emphasize that the geometry of the accretion flow
in the hard state is still not well constrained, and 
that observed spectral (including the relativistically broadened Fe K-$\alpha$ line) and spectro-polarimetric
signatures depend strongly on the plasma processes
responsible for energy dissipation in the plasma.
\end{abstract}
\section{Introduction}
Observations of Cyg X-1 have played a keyrole 
in driving the development of models to explain the X-ray emission from black hole X-ray binaries (BHXRBs) ever since the discovery of X-rays
from the source in 1964 \citep{1965Sci...147..394B}.
This includes the development of the standard model of geometrically thin, optically thick accretion disks \citep{1973A&A....24..337S,1973blho.conf..343N}.
The standard model posits that matter orbits the black 
hole on near-Keplerian orbits, locally dissipating the gravitational energy of the matter as it moves 
toward the black hole. 
\citet{1973A&A....24..337S} proposed that turbulence within the accreting gas provided the effective viscosity required for matter to sink toward the black hole, with the viscous stress and the pressure related by the $\alpha$ parameter. 
In the standard geometrically thin, optically thick accretion disks the magneto-rotational instability (MRI) driven by the differential rotation of the magnetized plasma \citep{RevModPhys.70.1} 
is believed to supply most of the viscosity. 
Although the vertical accretion disk structure depends 
on the microprocesses in the disk, the radial brightness temperature profile $T(r)$ is given by mass, energy, and angular momentum conservation alone \citep{1974ApJ...191..499P}. 
In the soft state of BHXRBs, the emission can be well described as diluted multi-temperature blackbody emission from the accretion disk atmosphere. Here, diluted means that the 
transport of radiation through the atmosphere with rarefied, hotter plasma in the upper layers results in blackbody-type emission with a temperature that exceeds the brightness temperature by the hardening factor of $\sim\,1.7$ \citep{1995ApJ...445..780S,2019ApJ...874...23D}. 
For \mbox{Cyg X-1}, the spectral energy distribution 
(SED) $E^2 dN/dE$ of the diluted blackbody emission 
peaks around 1\,keV. 

The earliest observations of Cyg X-1 already revealed evidence for a hard emission component \citep[e.g.,][and references therein]{1969A&A.....1...48R}. 
In the hard state, the power law index $\Gamma$ (with $dN/dE\propto E^{-\Gamma}$) 
has values between 1.5 and 2 \citep{2006A&A...447..245W,2010LNP...794...53B,2013A&A...554A..88G}.
The emission, commonly referred to as coronal emission, is attributed to accretion disk emission Comptonized by a 
hot plasma \citep{1976ApJ...204..187S,1976ApJ...206..910K,
1980A&A....86..121S,1979A&A....75..214P,1995ApJ...450..876T}.
For Cyg X-1, the plasma has a temperature of $k_{\rm B}\,T_{\rm e}\sim 100$\,keV and an optical depth $\tau \sim 1$ with a Compton $y$-parameter 
\[y\,=\, \frac{4 k_{\rm B} T_{\rm e}}{m_{\rm e} c^2}\,{\rm max}(\tau,\tau^2)\,\approx\,1
\]
with $k_{\rm B}$ being the Boltzmann constant, $m_{\rm e}$ the electron mass, and $c$ the speed of light.  
Photons of initial energy $\varepsilon_{\rm i}$ traversing the corona will exit the corona with 
an average energy of
$\varepsilon_{\rm f}\,=\,\varepsilon_{\rm i}\,e^y$ as long as 
$\varepsilon_{\rm i}<\varepsilon_{\rm f} \ll 4 k_{\rm b}\,T_{\rm e}$ 
\citep{Hurwitz1945,1957JETP....4..730K,1986rpa..book.....R}. 

The corona is frequently approximated 
in the lamppost approximation as a 
compact source of power law X-rays 
located on the spin axis of the black hole \citep[][]{1991A&A...247...25M}.
Alternatively, hot plasma, possibly structured, could be sandwiching 
the accretion disk \citep[the sandwich corona][]{1991ApJ...380L..51H,1993ApJ...411L..95H,1994ApJ...432L..95H} or 
could be located in the inner portion 
of a truncated accretion disk \citep{1997ApJ...482..400E}. 
The observations of broad emission lines, most prominently the Fe K-$\alpha$ emission line around 6.4\,keV, indicate that some of the coronal emission scatters off dense material close to the black hole
\citep[e.g.,][]{1989MNRAS.238..729F,2024ApJ...969...40D}. 
Gravitational and Doppler frequency shifts can explain the observed line shapes if the emission originates from a few gravitational radii $r_{\rm g}\,=\,GM/c^2$ (with $G$ being the gravitational constant and $M$ 
the black hole mass) from the black hole.
The coronal emission cuts off above $\sim$ 100\,keV \citep[e.g.,][]{2015MNRAS.451.4375F}.  

At higher energies, another component dominates
\citep{2017A&A...603A...8W}, possibly the hard energy tail from the plasma processes energizing the plasma close to the black hole \citep[e.g.,][]{2024PhRvL.132h5202G} or the emission from the base of the radio jet.   

The accretion of matter onto black holes involves a number of astrophysical processes. The black hole captures mass and magnetic field flux with rates of $\dot{M}$ and $\dot{\Phi}_{\bf B}$, respectively. 
The differential accretion flow in the Kerr spacetime is a highly nonlinear 
process involving plasma processes such as turbulence, the
MRI, magnetic field reconnection, and shocks.
Although thin disks have been simulated  \citep[e.g.,][]{2010MNRAS.408..752P,2010ApJ...711..959N}, 
a solid understanding of their actual structure is still missing. For example, \citet{2018ApJ...857...34Z} find that accretion disks threaded with a vertical magnetic field may accrete matter mostly through the magnetically dominated region  above and below the accretion disk (coronal accretion) and not, as previously thought, through the disk.
Magnetohydrodynamic simulations largely neglect plasma processes such as magnetic reconnection,
which may lead to the creation of new dynamically important components such as pair plasma.
The accretion flow converts gravitational 
energy, and possibly also rotational energy from the black hole \citep{1977MNRAS.179..433B}, into magnetic field, heat, bulk motion kinetic energy, radiation, and possibly into the masses of created pairs. 
The observer finally sees the radiation escaping the system, as well as some of the mechanical
energy going into winds and the jet.

This paper discusses the properties of the Cyg X-1 accretion flow in the soft state and in the hard state. We begin in Section\,\ref{obs} by reviewing the most pertinent observational constraints on these states, with a particular focus on recent X-ray polarization results from the {\it IXPE} \citep{2024mbhe.confE..31C} and 
{\it XL-Calibur} \citep{2021APh...12602529A} missions.
Section \ref{origin} discusses the theoretical implications of these observations, examining the sources and sinks of energy applicable to both states (Section\,\ref{s:conv}) and the energy flow into and out of the corona in the hard state (Section\,\ref{s:corona}).
In Section \ref{implications}, we discuss the implications of the X-ray polarization findings, centering on a novel model to explain the unexpectedly high X-ray polarization observed parallel to the radio jet in the soft state. Finally, Section\,\ref{disc} provides a summary and discussion of our results.
Throughout this paper, we give errors on the 68.27\% (1\,$\sigma$) 
confidence level. 
\section{Observational Constraints}
\label{obs}
\subsection{General Data and Constraints on the Mass Capture Rate}
Cyg X-1 is one of the most-observed objects in the sky, with a wealth of information across the electromagnetic energy spectrum \citep[see][for a recent review]{2024Galax..12...80J}.  The binary consists of a  $M\,=\,21.2\pm2.2\,M_{\odot}$ black hole
in a 5.599829(16) day orbit with an O-star
of mass 40.6$^{+7.7}_{-7.1}\,M_{\odot}$. The binary system is seen at an inclination of $i\,=\,27^{\circ}.51^{+0.77}_{-0.57}$
from its orbital axis \citep{2021Sci...371.1046M,2008ApJ...678.1237G}.  
The semi-major axis $a_{\rm bin}$ 
of 0.244 au is only 2.35 times larger than 
the radius $R_1$ of the companion 
star of 22.3$^{+1.8}_{-1.7}$
$R_{\odot}$. The eccentricity of the orbit is 0.0189$^{+0.0028}_{-0.0026}$.

\citet{2015A&A...576A.117G,2024A&A...691A..78L}
used X-ray observations to normalize 
models of the wind of Cyg X-1 companion 
star. They infer wind mass loss rates of 
$\dot{M}_{\rm wind}\,\sim\, 7\,\times\,10^{-6}\,M_{\odot}\,{\rm yr}^{-1}$.
For a wind velocity profile   
2100 km s$^{-1}$ $(1-R_1/r)^{\beta}$
with $\beta=1.5$, this implies 
a wind velocity of $v_{\rm wind}\,=\,916$\,km\,s$^{-1}$ at the location of Cyg X-1.
Adding this in quadrature to the orbital velocity of the black hole of $\sim$310\,km\,s$^{-1}$ gives a relative velocity
$v_{\rm rel}\,\approx\,$967\,km\,s$^{-1}$, and a 
Bondi-Hoyle capture radius \citep{1944MNRAS.104..273B} of:
\begin{eqnarray*}
R_{\rm cap}&=& \frac{2\, G\, M} 
{v_{\rm rel}^{\,\,2}}\\
&\approx&6\times10^{11}\,
\left(
\frac{M}{21.2\,M_{\odot}}
\right)
\left(
\frac{\rm 967\,km\, s^{-1}}{v_{\rm rel}}
\right)^{2}
{\rm cm}.
\end{eqnarray*}
The fraction of the wind mass captured 
by the black hole is thus:
\[
f_{\rm cap}\,\approx\,\frac{\pi R_{\rm cap}^{\,\,2}}{4 \pi a_{\rm bin}^{\,\,2}}\frac{v_{\rm rel}}{v_{\rm wind}}.
\]
The heating of the star by the X-rays from the accreting black hole and the gravitational 
pull from the black hole focus the accretion onto the black hole and amplify the accretion rate by a factor of $\xi\sim 3$ \citep{1982ApJ...261..293F},
giving a captured mass rate of:
\begin{eqnarray*}
\dot{M}_{\rm cap} & = & 
\xi\,f_{\rm cap}\,\dot{M}_{\rm wind}\\
& \approx & 
9.5\times 10^{18}
\left(\frac{\xi}{3}\right)\,
\left(\frac{\dot{M}_{\rm wind}}{7\,\times 10^{-6}\,M_{\odot} \,{\rm yr}^{-1}}\right)
 {\rm \,g\,s^{-1}}.
\end{eqnarray*}
The Eddington luminosity of Cyg X-1 is:
\[
L_{\rm Edd} = \frac{4 \pi\, G\, M\, c}
{0.2(1+X)}\,\approx 3\times 10^{39} {\rm \,erg \,s^{-1}}
\]
where we used the mean molecular weight per electron of $X\,=\,0.7$ for the wind 
of a hydrogen rich star \citep{2023pbse.book.....T}.
Combining the results, we infer:  
\begin{equation}
\dot{M}_{\rm cap}\approx 3.2\, L_{\rm Edd}/c^2.
\label{e:mdot}
\end{equation}
If Cyg X-1 converted 1\% of the mass energy from the captured stellar wind into radiation, it would thus allow it to shine with $\approx$3.2\% of the Eddington Luminosity - similar to the observed luminosities. 
Although the wind mass loss rate is uncertain 
by a factor of a few
\citep[see][]{2024A&A...691A..78L}, the result indicates 
that models with total radiative efficiencies 
well below 1\% are not viable.  
%
\subsection{Constraints from X-ray timing and spectral data}
\subsubsection*{Upper limits on the size of the emission region}
The most reliable upper limits of the 
size of the emission region come from the fast time variability of the X-ray fluxes.
As the regions causing large flares should be causally connected, flare durations of $\Delta t$ imply an upper limit on the source region size of:
\begin{equation}
R\,<\,\Delta t\,c.
\end{equation}
We neglect in the following the effects of gravitational time dilation and Doppler effects on the observed flux variability time scales as both effects 
are expected to impact the results by a few 10\% at most.   
We use the fast flares described by
\citet{2003MNRAS.343L..84G}. 
In the soft state the fastest flares 
exhibit exponential rise and decline 
times of $\sim$7\,ms which translate
into an upper limit on the size of 
the emission region of 
$\Delta R\,<\,$34\,$r_{\rm g}$.
A large fraction of the soft state emission is thought to come from between the innermost stable circular orbit (ISCO) and $5\,r_{\rm g}$
\citep[see][and Fig.\,\ref{f:eta}]{2014ApJ...790...29G}, indicating that the
disk collapses and replenishes with a sizable fraction of the speed of light. 
In the hard state, 
the fastest flares 
occurred on time scales of 27\,ms, corresponding to an upper limit on the size of the emission region of $\Delta R\,<\,$129\,$r_{\rm g}$. 
For the hard state, \citet{2010LNP...794...17G} reports a time lag between the 2-30 keV emission (presumably from the corona) and the Fe K-$\alpha$ emission around 6.4 keV (presumably from the 
reflection of the coronal photons by the accretion disk) of 15\,ms,
corresponding to $\Delta R\,<\,$72\,$r_{\rm g}$.
This limit is roughly consistent with the limit from the flux 
variability time scale, although it may constrain the 
distance of the corona from the reflecting accretion disk rather 
than the size of the corona.

Early Fe K-$\alpha$ line analyses indicated 
a very compact corona located within 
$5\,r_{\rm g}$ from the black hole \citep[e.g.,][and references
therein]{2015MNRAS.451.4375F}.
Later refined analyses weakened this 
constraint giving a lamp post height between 
19\,$r_{\rm g}$ and 36\,$r_{\rm g}$
 \citep{2018ApJ...855....3T}. 
These results will need to be revised 
once the location and properties 
of the corona are better constrained.

\subsubsection*{Luminosities and hard state stability}
\citet{2002ApJ...578..357Z,2003MNRAS.343L..84G}    estimate the bolometric luminosity of Cyg X-1 during the soft and hard states.  Correcting their results for the most recent Cyg X-1 mass and distance data \citep{2021Sci...371.1046M}, we 
infer time averaged luminosities of 2\%\,$L_{\rm Edd}$ in the soft state and 
0.5\%\,$L_{\rm Edd}$ in the hard state. The authors report flare luminosities of
12\%\,$L_{\rm Edd}$ in the soft state, and 10\%\,$L_{\rm Edd}$ in the hard state.

Cyg X-1 exhibits a remarkable stability in its hard state. During a $\sim$5-yr period between 1996 to 2002 in which it was in the hard state, the
3-12 keV photon index $\Gamma$ varied by typically less than $\Delta\Gamma \approx 0.1$, exhibiting a slow secular decline by $\Delta\Gamma \approx 0.2$ 
over the time period \citep{2003MNRAS.343L..84G}.

\subsection{X-ray polarization results}
{\it IXPE} measured the X-ray polarization of Cyg X-1 in the soft and hard states. The results were unexpected in several regards: the 
2-8\,keV polarization degrees (PDs) of (1.99$\pm$0.13)\% in the soft state \citep{2024ApJ...969L..30S} and (4.01$\pm$0.20)\% in the hard state \citep{2022Sci...378..650K,2025A&A...701A.115K} were higher than
expected when assuming that the black hole accretion disk is viewed at the
27$^{\circ}$.51 inclination of the binary. 
Furthermore, the polarization angle (PA) did not change much between states with PAs of 
-25$^{\circ}.7\pm 1^{\circ}.8$ in the soft state, and
-20$^{\circ}.7\pm 1^{\circ}.4$ in the hard state, respectively,
aligning with the radio jet within the accuracy with which the position angle
of the jet is known \citep{2001MNRAS.327.1273S}.

In the soft state, the optically thick emission from the accretion disk 
was expected to be polarized parallel to the accretion disk and perpendicular 
to the radio jet \citep{1960ratr.book.....C,1963trt..book.....S,1969ApJ...158..219A,1975MNRAS.171..457R,1976ApJ...203..701L,1977Natur.269..128C,2009ApJ...691..847L}.
Although Cyg X-1 always emits some power law emission even in the soft state,  \citet{2024ApJ...969L..30S} found that most models
of the combined thermal and power law emission 
predicted markedly lower
overall PDs than the observed ones, 
or PAs deviating from the observed ones.
The authors (including the two authors of this article) 
explained the polarization parallel to the radio jet by invoking an extremely high black hole spin parameter of $a \ge 0.96$ ($a\,=\,J_{\rm BH}\,/\,M\,c$ with $J_{\rm BH}$ being the black hole's angular momentum, and $-1\le a \le 1$). 
For such a high spin,
the {\tt kerrC} code \citep{2022ApJ...934....4K}
predicts that returning accretion disk and coronal emission dominate the overall polarization and give a net polarization 
perpendicular to the accretion disk \citep{2009ApJ...701.1175S}.

The PD of the hard state roughly agreed with the expectations for a corona extended laterally parallel to the accretion disk \citep{1985A&A...143..374S,1993A&A...275..337P,
2010ApJ...712..908S}, but required the inner accretion disk to be seen at inclinations of $i\ge 40^{\circ}$, 
higher than the orbital inclination of $27^{\circ}.51$ \citep{2022Sci...378..650K}, 
or the corona moving at $>$40\% 
of the speed of light parallel 
to the jet \citep{1998ApJ...496L.105B,1999ApJ...510L.123B,2023ApJ...949L..10P}.

From 15 to 60\,keV {\it XL-Calibur} hard state observations show a continuation of the rather low PD with a PA parallel to the radio jet. This is consistent with the {\it IXPE} and {\it XL-Calibur} emission being dominated by the same emission process in the hard state \citep{2025arXiv250723126A}.

The Comptonized emission cuts off between 100 and 200\,keV and gives way to another power-law component emitted by non-thermal high-energy particles accelerated in the corona or further away from the black hole in the jet \citep[e.g.,][and references therein]{Frontera_2001,2003MNRAS.343L..84G,2006A&A...446..591C,2021MNRAS.500.2112K}.
The results from {\it AstroSAT} and {\it INTEGRAL} indicate that the PD may indeed increase drastically and the PA may swing above 100\,keV \citep{2011Sci...332..438L,2012ApJ...761...27J,2015ApJ...807...17R,2024ApJ...960L...2C}.
We anticipate that COSI will be able to measure the $>$200\,keV polarization properties with smaller
systematic errors \citep{2022hxga.book...73T}.
\section{Origin of the X-ray emission}
\label{origin}
\subsection{Conversion of the gravitational energy of the accreted material and the rotational energy of the black hole into the observed luminosity}
\label{s:conv}
The following discussion uses the Kerr metric in Boyer Lindquist coordinates 
$x^{\mu}\,=\,(t,r,\theta,\phi)$ \citep{1967JMP.....8..265B}.
The energy at infinity (including rest mass energy) of a mass $m$ in an equatorial Keplerian orbit is:
\begin{equation}
E(a,r)\,=\,\frac{r^{3/2} - 2\,r^{1/2} + a}{r^{3/4} \sqrt{r^{3/2} - 3\,r^{1/2} + 2\,a}}
\label{e:eta2}
\end{equation}
in units of $m\,c^2$.  
Here, $r$ is given in units of $r_{\rm g}$, 
and {\tt sign}  is the sign function \citep{1972ApJ...178..347B}.
The function $E(a,r)$ is 1 for 
$r\rightarrow \infty$, and decreases as 
the mass loses gravitational energy 
by moving to orbits closer to the black hole. 

Matter moving on quasi-Keplerian orbits from $r_2$ to $r_1$ generates 
the luminosity:
\begin{equation}
L_{\rm grav}(a,r_1,r_2)\,=\,\eta_{\rm grav}(a,r_1,r_2)\,\dot{M}\,c^2
\label{e:L}
\end{equation}
with
\begin{equation}
\eta_{\rm grav}(a,r_1,r_2)\,=\,E(a,r_2)-E(a,r_1).
\label{e:eta1}
\end{equation}
being the efficiency of converting 
accreted mass energy into free energy.
The black hole spin may provide additional power via 
the Blandford-Znajek (BZ) process 
\citep[][and references therein]{1977MNRAS.179..433B,2009JKPS...54.2503K}. 
The BZ luminosity is given by:
\begin{equation}
L_{\text{BZ}}\, =\, \frac{\kappa}{4\pi c} \, \Omega_{\text{H}}^2 \, \Phi_{\text{BH}}^2 \, f(\Omega_{\text{H}})
\end{equation}
with $\kappa\sim 1/20$ depending on the magnetic field configuration close to the black hole,
$\Omega_{\rm H}\,=\,a\,c\,/\,2\,r_{\rm H}$ being 
the angular frequency at the event horizon at radial coordinate $r_{\rm H}$,
$\Phi_{\text{BH}}$ the magnetic flux threading the event horizon, and
$f(\Omega_{\text{H}})\,\approx\,1$
for $a<0.95$, i.e., for all 
but the most rapidly 
spinning black holes
\citep[][]{2011MNRAS.418L..79T}.
The BZ luminosity can be parameterized as:
\begin{equation}
    L_{\text{BZ}}\,=\,\eta_{\rm BZ} \dot{M}\,c^2
\end{equation}
where $\eta_{\rm BZ}$ 
may be of order unity for accretion flow configurations
that support large  $\Phi_{\text{BH}}$ \citep[][]{2011MNRAS.418L..79T}. 

The mass accretion and BZ luminosities power the observed emission ($L_{\rm rad}$), 
winds and jets ($L_{\rm wind/jet}$), and other forms of energy that evade detection (e.g., energy converted into the rest mass of pairs, and energy carried by undetected hot or magnetized plasma):
\begin{equation}
L_{\rm grav}+L_{\rm BZ}\,=\,L_{\rm rad}+L_{\rm wind/jet}+L_{\rm other}.
\end{equation}

Neglecting $L_{\rm BZ}$ for the time being, we use the observed radiative luminosity to infer a minimum mass accretion rate via:
\begin{equation}
\eta_{\rm rad}\,L_{\rm grav}\,=\,
\eta_{\rm rad}\,\eta_{\rm grav} \dot{M} \,c^2 \,\ge\,
    L_{\rm rad}
    \end{equation}
with $\eta_{\rm rad}\,\approx\,\frac{1}{6}$.
The latter fraction results from the product of two factors.  
We estimate that roughly one-third of the free 
energy goes into the magnetic field,
given that non-radiative simulations of the MRI  
indicate equipartition 
between the thermal plasma energy 
and the magnetic field \citep[e.g.,][and references therein]{2022A&A...659A..91W}. 
Rather than using the factor one-half, we use the factor one-third as part of the energy will be converted into radiation.  We estimate furthermore that one-half of the magnetic energy can be converted into radiation as shown by particle in cell (PIC) simulations \citep[see][and references therein]{10.1093/mnrasl/sly157,2025ARA&A..63..127S}.

\begin{figure}
  \centering
  \includegraphics[width=0.45\textwidth]{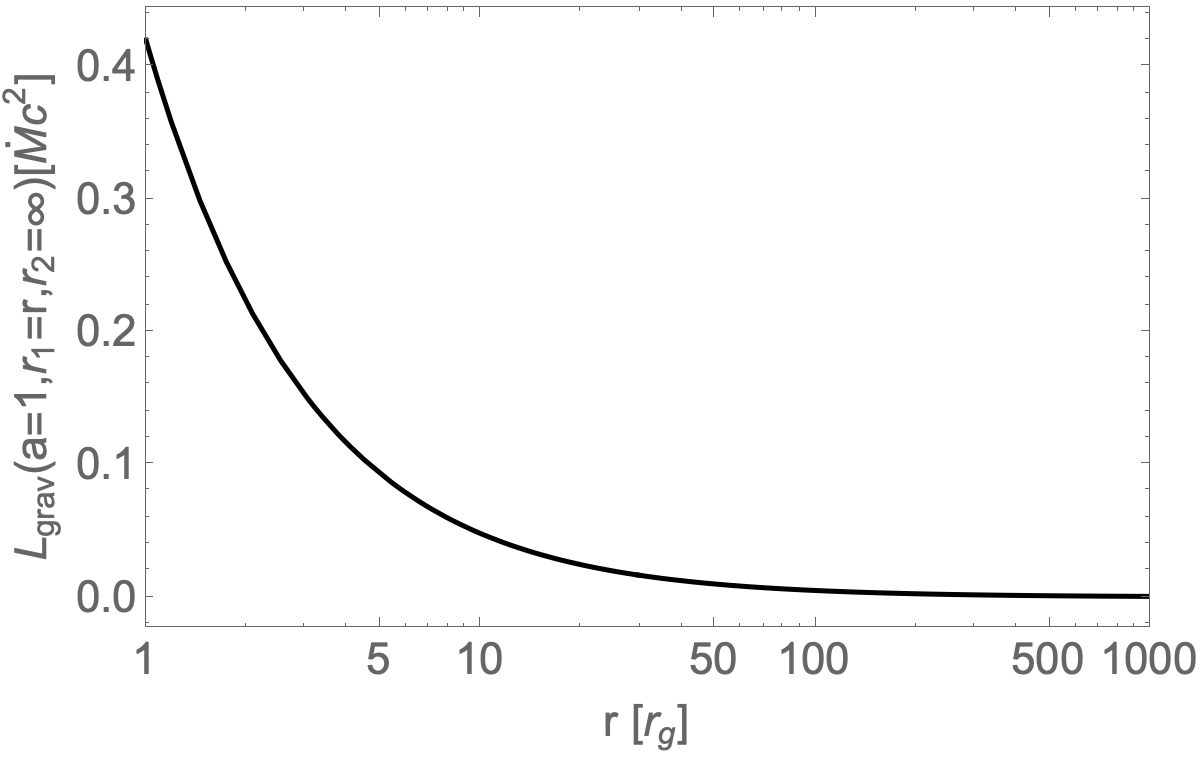}
  \caption{Efficiency $\eta_{\rm grav}(r_1=r,r_2=\infty)$ between the conversion of gravitational energy to free energy when mass moves on Keplerian orbits from $r\rightarrow\infty$ to $r$.  
  }
  \label{f:eta}
\end{figure}
Figure \ref{f:eta} shows $\eta_{\rm grav}(a,r_1,r_2=\infty)$ for a maximally spinning 
black hole ($a\,=\,1$). In the soft state, matter is believed to sink all the way to the innermost stable orbit, allowing for a high conversion efficiency $\eta_{\rm grav}$ between $\dot{M}\,c^2$ and $L$ up to 42.3\%.

We consider two corona models for explaining the hard state: 
a compact corona that draws its energy from the region 
$r_{\rm ISCO}\,<\,r\,<\,5\,r_{\rm g}$
of a black hole with $a\,=\,0.998$ ($r_{\rm ISCO}\,=\,1.237\,r_{\rm g}$) and a more extended corona that 
draws its energy from the region  $10\,r_{\rm g}\,<\,r\,<\,130\,r_{\rm g}$ with the spin $a$ being unconstrained. 
The choice of the particular two scenarios is somewhat arbitrary, 
but the basic idea is to distinguish between a compact corona close 
to the black hole and a more extended region 
just satisfying the size limits from flux variability. 

The compact corona gives us 
$\eta_{\rm grav}$ of 22.72\% and
$\eta_{\rm tot}\,=\,\eta_{\rm grav}/6$ of 3.79\%, 
requiring mass accretion rates of 
between 0.13\,$L_{\rm Edd}/c^2$ and 2.64\,$L_{\rm Edd}/c^2$ to explain 
the average hard state luminosity of
0.5\% Eddington and the flare luminosity of 
10\% Eddington, respectively.  
The extended corona gives us 
$\eta_{\rm grav}$ of 4.43\% and
$\eta_{\rm tot}\,=\,$ of 0.74\%
and requires mass accretion rates of between 
0.68\,$L_{\rm Edd}/c^2$ and 13.56\,$L_{\rm Edd}/c^2$ for the average and flare luminosities.
The compact corona thus requires smaller mass 
accretion rates more in line 
with the estimate from Equation (\ref{e:mdot}).
We summarize the values of $\eta_{\rm grav}$ and 
the implied accretion rates in Table\,\ref{t:p}.

In the above analysis of the soft state and the hard state energetics, we neglected two effects. First, additional luminosity can originate from the plunging region between the ISCO and the black hole horizon. 
For a thin disk extending from $r_{\rm ISCO}$ to infinity, the luminosity from the plunging region is estimated to increase the total luminosity by 10\% \citep[see e.g.,][]{2010MNRAS.408..752P,2010ApJ...711..959N,2022MNRAS.515..775H,2024MNRAS.531..366M,2024MNRAS.533L..83M}.
Furthermore, our analysis does not account for the fact that a large corona extending from $r_1$ to $r_2$ may draw its energy from the inner accretion flow at $r\,<r_1$. This possibility could make a spatially extended corona more efficient.

As mentioned above, we assume one-third of the gravitational energy of the accreting material is converted into magnetic fields that power the corona. The remaining two-thirds could still power the accretion disk that reaches into or through the corona. If there is an accretion disk inside the corona, it would be less luminous than predicted by the standard thin disk theory, which assumes that 100\% 
of the gravitational energy is locally radiated away. 
Another configuration could be that the flow at small distances from the black hole is not a continuous accretion disk but is made of cold clumps that sink toward the black hole \citep{2022ApJ...935L...1L}, which can inject soft photons into the corona and can reflect coronal emission.
\subsection{Energy transport into and out of the corona}
\label{s:corona}
\begin{figure}
  \centering
  \includegraphics[width=0.45\textwidth]{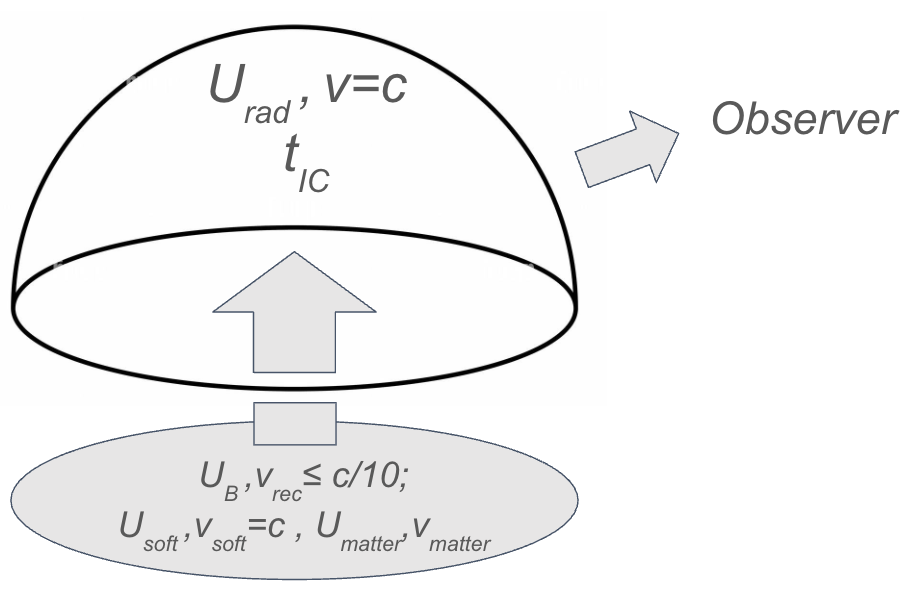}
  \caption{The observed X-ray fluxes set a lower limit on the energy density $U_{\rm rad}$ of
  X-rays in the corona. 
  The energy has to be supplied by magnetized plasma with the 
  energy density $U_{\rm B}$, 
  by soft radiation with energy density
  $U_{\rm soft}$, or 
  bulk motion of plasma with 
  energy density $U_{\rm matter}$ 
  streaming into the corona, each with its characteristic velocity. 
  }
  \label{f:sketch}
\end{figure}
This section discusses the energy flows into and out of the corona for the hard state, focusing on the two corona models from the previous section: 
compact coronal emission volumes in the inner $5\,r_{\rm g}$ region 
of the accretion flow, and a more extended coronal region extending from
$10\,r_{\rm g}$ to $130\,r_{\rm g}$.
Following the approach of
\citet{1979ApJ...229..318G,1994ApJ...432L..95H}, and
\citet{2017ApJ...850..141B}
(called AB17 in the following), we analyze the flows of energy in its various forms, radiation, ions and electrons, and magnetic field into and out of the corona accounting for the flow velocities of these components
(see Fig.\,\ref{f:sketch}), and the time scales of the conversion from one form of energy into another. 
We largely use the nomenclature of AB17.

The observed emission is likely the accretion disk emission
Comptonized by cold or hot plasma 
and some non-thermal particles \citep[AB17,][]{2020ApJ...899...52S,
2021MNRAS.507.5625S,
2023MNRAS.518.1301S,
2024MNRAS.527.6065G,
2024PhRvL.132h5202G}.
The hot and non-thermal particles 
cool radiatively
\citep[e.g.,][]{1988ApJ...334L...5G,2006A&A...451..739K,2009MNRAS.392..570M,2009ApJ...690L..97P}, possibly creating
$\gamma$-rays which can pair-produce
and modify the properties of the
upstream and downstream plasma
\citep[e.g.,][]{2021MNRAS.508.4532M,2024MNRAS.52711587M}.
Given the bolometric luminosity $L$ and the radius $R$ of the emission region (assumed to be a hemisphere), the radiation energy density inside the emission region is \citep{1986rpa..book.....R}:
\begin{equation}
U_{\rm rad}\,=\,\frac{L}{4\pi R^2 c}.
\label{e:urad}
\end{equation}
The Compton compactness parameter \citep{2009ApJ...690L..97P} is: 
\begin{equation}
l_{\rm rad}\,=\, \frac{U_{\rm rad}\,\sigma_{\rm T}\, R}{m_{\rm e} c^2}
\end{equation}
The Compton cooling time 
of electrons with velocities $\beta_{\rm e}\,c$ and
Lorentz factors 
$\gamma_{\rm e}\,=\,\left(1-\beta_{\rm e}^{\,2}\right)^{-1/2}$ 
is given by:
\begin{equation}
t_{\rm IC}\,=\,\gamma_{\rm e}/\dot{\gamma_{\rm e}}\,=\,\frac{3}{4}\frac{1}{l_{\rm rad}\,\beta_{\rm e}^{\,2}\,\gamma_{\rm e}}\,\frac{R}{c}.
\end{equation}

If the corona is powered by  magnetized plasma moving from the disk into the corona  
with the velocity $v_{\rm rec}$ 
from radius $r_1$ to $r_2$ with $r_2\,\approx\,R$, we infer an average upstream magnetic field energy density of:
\begin{equation}
<\!U_{\rm B}\!>\,=\,\frac{L}{\pi\,(r_2^{\,2}-r_1^{\,2})\,v_{\rm rec}}.
\label{e:ub1}
\end{equation}
The expression accounts for two factors of 2 which cancel each other: the disk provides magnetic flux into both hemispheres (doubling the area in the denominator), but magnetic reconnection converts only  $\sim$50\% of $U_{\rm B}$ into bulk and random particle kinetic energy (doubling the required energy density). For fast reconnection in the collisionless regime, 
$v_{\rm rec}\,\sim\,c/10$
\citep[e.g., ][]{2008ApJ...688..555G,2016ASSL..427..473U,2025ARA&A..63..127S}. 

We can compare the magnetic field energy density 
$<\!U_{\rm B}\!>$ required to power the corona with 
the magnetic field energy density in the accretion disk $<\!U_{\rm B,disk}\!>$ averaged over the area of the disk from which the corona draws its energy.
In a steady state situation, the angular momentum transported by the accreting matter toward the black hole $\dot{J}\,=\,\dot{M}\sqrt{G\,M\,r}$ is balanced by the angular momentum transport by the turbulent shear stress $\dot{J}\,=\,2\,\pi\,r\,(2\,H)\,\tau_{\rm r\phi}$.
Here, $H\,\approx\, h\,r$ is the accretion disk thickness at radial distance $r$ 
and $\tau_{\rm r\phi}\,=\,\alpha\,P$ 
is the shear stress. 
Shakura and Sunyaev's $\alpha$ parameter is expected to have values between 0.01 and 0.1, and $P$ is the total pressure. 
Combining these equations gives:
\begin{equation}
\alpha\,P\,=\,\frac{\dot{M}\sqrt{G\,M\,r}}{4\pi\,r^2\,H}.
\end{equation}
We estimate that the magnetic field carries 1/3rd of that pressure, so that $U_{\rm B}(r)\,=\,P(r)/3$.
Averaging over disk area between $r_1$ and $r_2$, this gives:
\begin{equation}
    <\!U_{\rm B,disk}\!>\,=\,\frac{\int_{r_1}^{r_2}\,dr\,2\pi r\,(P/3)}{\pi(r_2^{\,2}-r_1^{\,2})}.
    \label{e:ub2}
\end{equation}
We use this equation with the mass capture rates 
mentioned in the previous sections 
and with $\alpha\,=\,0.1$ and $H\,=\,r/10$.

Using our estimates or limits on the 
radiation energy density $U_{\rm rad}$ from Equation\,(\ref{e:urad}) and the magnetic field energy 
density $<\!\!U_{\rm B}\!\!>$ from Equation\,(\ref{e:ub1}), we estimate the maximum 
Lorentz factors of electrons (or positrons) accelerated by reconnection in the corona.
Three dimensional PIC simulations show that the maximum Lorentz factor is given by the classical burnoff limit at which energy gains in the reconnection electric field equal the radiative energy losses  per unit time \citep[e.g.,][]{2020ApJ...899...52S,2025ARA&A..63..127S}:
\begin{equation}
    \gamma_{\rm max}\,=\,\sqrt{\frac{e\,v_{\rm rec}\,B_{\rm rec}}{(4/3)\,\sigma_{\rm T}\,U}}.
\end{equation}
Here, $e$ is the electron charge. We assume 
that the reconnection magnetic field is $B_{\rm rec}\,=\,\sqrt{8\pi\!<\!\! U_{\rm B}\!\!>}$, and $U\,=\,U_{\rm rad}$ for Compton cooling and $U\,=\,<\!\!U_{\rm B}\!\!>$ for synchrotron cooling.
Note that the synchrotron emission may be suppressed because it may be self-absorbed (AB17).
The estimates of $\gamma_{\rm max}$ can be used to infer if the accelerated electrons have enough 
energy to create pairs via 
$\gamma\gamma$-pair production.

The optical depth of the corona $\tau\,\sim\,1$ implies the electron 
(and positron) density of: 
\begin{equation}
n_{\rm e}\,=\,\frac{\tau}{R\,\sigma_{\rm T}}.
\label{e:ne}
\end{equation}
Note that this would underestimate the true $n_{\rm e}$ as it does not account for 
the reduction of the optical depth 
owing to the somewhat parallel motion of the electrons and photons.

We can combine $n_{\rm e}$ with the $B$-field estimate of Equation\,(\ref{e:ub1}) to estimate the magnetization of a pair plasma (subscript e) or an electron-proton plasma (subscript p, assuming $n_{\rm p}\,=\,n_{\rm e}$)
supplying the energy to the corona:
\begin{equation}
\sigma_{\rm e/p}\,=\,
\frac{2\,U_{\rm B}} {n_{\rm e}\, m_{\rm e/p} \, c^2}
\end{equation}
The X-ray polarization results indicate that the Comptonizing plasma moves with velocities of $\beta_{\rm bulk}\sim c/2$. The bulk kinetic energy  
of pair plasma $E_{\rm e^+e^-}$
between $r_1$ and $r_2$ is approximately
\begin{equation}
E_{\rm e^+e^-}\,=\,(\gamma_{\rm bulk}-1)\frac{4\pi}{3}(r_2^{\,\,3}-r_1^{\,\,3})\,n_{\rm e}\,m_{\rm e}\,c^2
\label{e:epem}
\end{equation}
with $\gamma_{\rm bulk}\,=\,(1-\beta_{\rm bulk}^{\,2})^{-1/2}$.
If the pair plasma escapes on a time scale of 
$t_{\rm esc}\,=\,(r_2-r_1)/(\beta c)$, we infer that the luminosity 
\begin{equation}
L_{\rm e^+e^-}\,=\,E_{\rm e^+e^-}/t_{\rm esc}
\end{equation}
is required to continuously replenish it.
We get the corresponding equations for a 
Comptonizing electron-ion plasma 
(subscript eI) by replacing 
$m_{\rm e}$ in Equation (\ref{e:epem}) 
with the mean molecular 
mass per electron  of 1.3 $m_{\rm p}$.

\begin{deluxetable*}{lp{5cm}ccccc}
\tablenum{1}
\tablecaption{\label{t:p} Parameter constraints on a compact corona and a spatially extended corona.}
\tablehead{
\colhead{Symbol} &\colhead{Name} & \multicolumn{2}{c}{Compact Corona} & 
\multicolumn{2}{c}{Extended Corona} & \colhead{Units}\\
& & 0.5\%\,$L_{\rm Edd}$& 10\%\,$L_{\rm Edd}$&
0.5\%\,$L_{\rm Edd}$& 10\%\,$L_{\rm Edd}$& 
}
\startdata
$r_1,r_2$ & Radial range of energy extraction    & \multicolumn{2}{c}{1.237$-$5} & \multicolumn{2}{c}{10$-$130} &$r_{\rm g}$\\
$\eta_{\rm grav}$ & Mass-to-energy conv. efficiency & \multicolumn{2}{c}{22.72\%} & \multicolumn{2}{c}{4.43\%} &$r_{\rm g}$ \\ \hline
$\dot{M}_{\rm cap}$ & Lower limit on captured mass & 0.13&2.64 &0.68 &13.56&$L_{\rm Edd}/c^2$\\ 
$\dot{M}_{\rm cap}$ & Lower limit on captured mass & $4.58\times10^{17}$&$9.16\times 10^{18}$ &$2.35\times10^{18}$&$4.70\times 10^{19}$ &g s$^{-1}$\\ 
$<\!U_{\rm B,disk}\!>$ & Available disk B-field energy dens. 
&$2.86\times10^{14}$ &$5.72\times 10^{15}$ & $1.04\times 10^{12}$ & $2.08\times*10^{13}$& erg cm$^{-3}$\\  \hline
%
$U_{\rm rad}$ & Required Rad. energy density & 1.70$\times$10$^{11}$& $3.40 \times 10^{12}$&$2.51 \times 10^{8}$ &$5.02 \times 10^{9}$&erg cm$^{-3}$\\
$l_{\rm rad}$ & Compton compactness & 2.16 & 43.2& 0.083&1.66& -\\ 
$t_{\rm IC}$ & Compton cooling time ($\gamma_{\rm e}=1$)& 0.35&0.017 &9.03&0.451& $R\,/\,c$ \\ 
$<\!U_{\rm B}\!>$ & Required B-field energy density& 7.24$\times 10^{12}$&$1.45\times 10^{14}$ &$1.01\times 10^{10}$&$2.02\times 10^{11}$& erg cm$^{-3}$ \\ 
$\gamma_{\rm max,rad}$ & Compton burnoff Lorentz factor & 6.56$\times 10^4$ & 3.10$\times 10^4$ & 3.30$\times 10^5$ & 1.56$\times 10^5$& -\\
$\gamma_{\rm max,B}$ & Synchrotron burnoff Lorentz factor& 1.01$\times 10^4$ & 4.75$\times 10^3$ & 5.02$\times 10^4$ & 2.46$\times 10^4$& -\\
\hline
$n_{\rm e}$ & Electron (and positron) density &  
\multicolumn{2}{c}{$9.60\times10^{16}$}
&\multicolumn{2}{c}{$3.69\times10^{15}$}& cm$^{-3}$ \\ 
$\sigma_{\rm e}$ & Electron magnetization & 184 & 3682&6.69&134& - \\ 
$\sigma_{\rm p}$ & Proton magnetization  & 0.10 & 2.01&0.0036&0.073& - \\ 
$L_{\rm e+e-}$&Power to launch e$^+$e$^-$ wind &\multicolumn{2}{c}{7.81$\times 10^{-5}$}
&\multicolumn{2}{c}{1.68$\times 10^{-3}$}&$L_{\rm Edd}$\\
$L_{\rm eI}$&Power to launch electron-ion wind
&\multicolumn{2}{c}{0.19}&\multicolumn{2}{c}{4.01}&$L_{\rm Edd}$\\
\hline
\enddata
\end{deluxetable*}

Table \ref{t:p} gives the inferred parameters for the average and flare hard state luminosities for the two extreme corona models. 
The numbers derived for the average luminosity of 0.5\% $L_{\rm Edd}$ are more robust than those for the flare luminosity of 10\% $L_{\rm Edd}$ as the latter may correspond to rare fast discharges
of energy accumulated over longer times.
Whereas the compact corona has a Compton compactness parameter $l_{\rm rad}\,=\,2.16$, the extended corona is not 
compact with $l_{\rm rad}\,=\,0.083$.
The Compton cooling time $t_{\rm IC}$ is is only 
0.35\,$R/c$ for the compact corona but 9.03\,$R/c$ for the extended corona.
Thus, the heat of plasma streaming into the corona cannot provide the energy required for explaining 
the observed emission of the compact corona 
\citep[as emphasized by][]{1979ApJ...229..318G}, 
but can do so for the extended corona.
Whereas the compact corona requires that energy 
be transported via magnetized plasma, plasma bulk motion 
or radiation, the extended corona may be powered 
by hot plasma. 

For the compact and extended coronas, the average magnetic 
field energy densities required for powering the corona 
through magnetic field reconnection are $40-100$ times 
smaller than the area-averaged magnetic field densities in the disk, consistent with a disk having a higher magnetic 
field energy density than the matter above the disk.

The maximum electron burnoff Lorentz factors
are on the order of $10^4$, high enough to
emit sufficiently high energy synchrotron photons or
inverse Compton photons by scattering photons from the accretion disk to create pairs in 
photon pair production processes.
The generation of pair plasma with 
an optical depth of $\sim$1 may 
be a self-regulating process 
that explains 
the remarkable stability of the spectral properties 
of Cyg X-1 in the hard state mentioned above (AB17).

We infer electron magnetizations 
of 184 and 6.69 for the cases of the compact and extended coronas, respectively.
The inferred plasma magnetizations 
are much lower if the plasma includes 
protons.
Even for the extended corona, we cannot
exclude that the plasma produces pairs, as 
small regions may be magnetized much more strongly than average.  

Whereas the luminosities to accelerate 
coronal pair plasma to $\sim c/2$ are 
small compared to the Eddington luminosity
for the compact and extended coronas, 
the acceleration of electron-ion plasma
requires 19\% of $L_{\rm Edd}$ for the compact corona and $\sim 4 \times L_{\rm Edd}$ for the extended corona.
Supplying this luminosity at 
$\eta_{\rm grav}\,\approx 4.43\%$
would require a prohibitive 
mass accretion rate.
We can thus exclude the possibility of an 
extended electron-ion corona 
outflowing with $\sim c/2$.

If the BZ effect contributes to powering the corona, it would do so most likely
at small distances from the black hole spin axis \citep{2019MNRAS.487.4114Y,2025A&A...701A..62M}.
The BZ thus seems to be more likely to play a role in the 
compact corona scenario, bolstering the argument that the compact corona can power the observed emission more readily than the extended corona.
%
\section{Implications of the X-ray polarization results}
\label{implications}
The {\it IXPE} Cyg X-1 observations revealed strong polarization parallel to the radio jet in the soft state and in the hard state. As mentioned above, explaining the results with the standard model requires a very high black hole spin for the soft state,  and either high inclinations or a mildly relativistically outflowing corona for the hard state.

In this section, we discuss a model to explain the soft state polarization, invoking a layer of electron positron pairs at the Compton temperature of the accretion disk, moving away from the disk with $c/2$. 
Such a pair layer could form following the acceleration of electrons in magnetic reconnection, leading to the emission of inverse Compton $\gamma$-rays and photo-pair-production processes. 
The pairs would likely accelerate to mildly relativistic velocities owing to two mechanisms.
{\bf (A) The Compton rocket:} The hot plasma cooling through inverse Compton scattering of the anisotropic radiation field from the accretion disk will accelerate away from the accretion disk, converting its random motion 
into directed motion. The effect was found to give 
rise to moderate outflow velocities   
\citep{1981ApJ...243L.147O,1981ApJ...251L..49C,1982MNRAS.198.1109P} in the case of electron-ion plasmas that cool on an anisotropic radiation field. 
The terminal bulk Lorentz factors will be higher if 
(i) the plasma is a pair plasma (e.g. formed by 
magnetic reconnection) with negligible ion loading or 
(ii) reconnection continues to heat the plasma while it accelerates.
{\bf (B) Radiation pressure on cold pairs:} even a cold pair plasma will naturally acquire velocities
around $\sim$50\% of the speed of light when exposed to an anisotropic radiation field  \citep[][called AB98 in the following]{1998ApJ...496L.105B} as well as \citep[][]{1989A&A...216..294I,1996ApJ...458..514L,1999MNRAS.305..181B,1999ApJ...510L.123B}.
The e$^+$ and e$^-$ accelerate to a velocity at which the scatterings do not lead to a momentum exchange anymore.
For optically thin plasmas, the terminal velocity 
depends on the anisotropy of the radiation field with the 
emission from a limb darkened scattering atmosphere 
giving a velocity of $\beta\,=\,0.52$.
If the electron fluid is optically thick, a velocity 
profile is established with velocities rising 
throughout the accelerating and expanding pair plasma 
as the flow ``straightens itself out'' and the pair and photon momenta get increasingly aligned down the flow.
AB98 shows that a pair plasma with an optical depth of 3 
will acquire a velocity profile with velocities increasing from $\beta\,=\,0.36$ to $\beta\,=\,0.705$ as the photon
wavevectors and the pair momentum vectors align.

AB98 shows that the Comptonization in such mildly relativistic outflows leads to a strong polarization parallel to the outflow velocity and invokes the effect 
to explain the polarization of the optical emission from active galactic nuclei (AGN) parallel to the direction of their radio outflows. \citet{2023ApJ...949L..10P} invoked an outflowing corona to explain the strong polarization of the Cyg X-1 hard state emission.

In the case of the Cyg X-1 discussed here, the pair layer could be created 
as a consequence of  magnetic reconnection or turbulence driven by the Keplerian shear stresses or
by stresses from the accretion disk torquing the tenuous plasma above the accretion disk.
Furthermore, dissipation of BZ Poynting flux could play a role. Once created, the pairs accelerate owing to the Compton rocket effect and the radiation pressure.

In the standard thin disk model, the accretion disk atmosphere hardens the energy spectrum from the accretion disk by the hardening factor of $\sim$1.7
\citep{1995ApJ...445..780S,2019ApJ...874...23D}.
This result follows for stationary electron-ion accretion disk atmospheres that are kept in place by the gravity 
of the ions.  In a variation of the standard model, \citet{2024ApJ...962..101Z,2024ApJ...967L...9Z} discuss that a warm (but not outflowing) optically thick Comptonization layer would impact the X-ray energy spectra and the inferred black hole spin estimates. 
In the scenario proposed here, the standard electron ion atmosphere may still be present, but a Compton thick layer of pairs is added. The pairs, unimpeded by the weight and inertia of ions, accelerate owing to the 
Compton rocket and radiation pressure effects. 

The polarization of the X-rays in both the soft state and the hard state might thus be affected by Comptonization 
in a mildly relativistically moving outflow. 
In the soft state the pair plasma has the Compton temperature 
\citep{1981ApJ...249..422K}
of the possibly diluted blackbody emission of the underlying geometrically thin, optically thick accretion disk emission. In the hard state, the pair plasma gains additional 
internal energy, i.e.,  bulk plasmoid motion and/or 
motion on all scales from turbulence. 
The state transition could be caused by reconnection happening in a different regime 
\citep{2008ApJ...688..555G} as a consequence 
of a reconfiguration of the accretion flow.

In the following we present results from radiation transport calculations for photons passing through  plane parallel pair atmospheres with certain $\beta$-profiles. Whereas the calculations of AB98 focused on the optical emission from AGNs where all scatterings occur in the Thomson regime and electrons and photons exchange negligible amounts of energy,
we focus here on the X-ray emission.
In this case, electrons and photons exchange energy, and the results become energy dependent. Our code uses the Comptonization engine from the {\tt kerrC} code. %
Following AB98, we characterize the thickness of the pair atmosphere with the vertical thickness $t_{\rm z_{\rm max}}\,=\,\int_0^{z_{\rm max}}\,n_{\rm e}(z)\,\sigma_{\rm T}\,dz$ (all quantities in the stationary reference frame).
We inject blackbody emission from a plasma at temperature $T_{\rm i}$ with the angular distribution and polarization 
given by Chandrasekhar's classical result for
an optically thick scattering atmosphere \citep{1960ratr.book.....C}.
The Lorentz invariant optical depth $\tau$ defined by 
$d\tau\,=\,n_{\rm e}^*\,\sigma_{\rm T}\,ds^*$ (all starred quantities in the plasma rest frame) is used to decide
if a scattering occurs. Note that $ds^*\,=\,\gamma\,(1-\beta \mu)\,ds$, $n^*\,=\,(1/\gamma)\,n$, so that 
$d\tau\,=\,(1-\beta\mu)\,n_{\rm e}\,\sigma_{\rm T}\,ds$
$=\,(1-\beta\mu)d t_{\rm z} / \mu$.
If the photon scatters, the photon four wavevector 
$k^{\mu}$ and four polarization vector $f^{\mu}$ are
transformed into the plasma frame. 
A Lorentz factor is drawn according to a thermal 
distribution with a temperature $T_{\rm p}$ and with
direction cosines $\mu_{\rm \gamma e}$ distributed
according to the probability distribution 
$p(\mu_{\rm \gamma e})\,\propto\,(1-\beta_{\rm e}\mu_{\rm \gamma e})$ with $\mu_{\rm \gamma e}$ being the 
cosine between the photon and electron directions.
The scattering is simulated in the rest frame of the electron, making use of the fully relativistic Fano scattering matrix that includes the effect of the Klein Nishina scattering cross section.
After the scattering, the photon wavevector and the polarization vector are transformed back, 
first into the plasma frame, then into the stationary frame.
Photons are tracked until they leave the atmosphere 
at $t_{\rm z}\,=\,0$ or $t_{\rm z}\,=\,t_{\rm z_{\rm max}}$.
The photons reaching $t_{\rm z}\,=\,0$ are discarded and the ones reaching $t_{\rm z_{\rm max}}$ are sorted into inclination bins, where their flux and Stokes $Q$ and $U$ energy spectra are acquired. 
We use the classical convention \citep[e.g.,][]{1960ratr.book.....C,1985A&A...143..374S} that $Q<0$ for PAs 
parallel to the surface normal of the atmosphere, 
and $Q>0$ for PAs parallel to the atmosphere. 
Note that owing to the symmetry of the problem, 
Stokes $U$ is zero, and the PD is simply given by $|Q|/I$. We use the code with small optical depths and  constant $\beta$-values, and for the 
$t_{\rm z}\,=\,3$ $\beta$-profile in Figure\,3 of AB98. 

\begin{figure}
  \centering
  \includegraphics[width=0.45\textwidth]{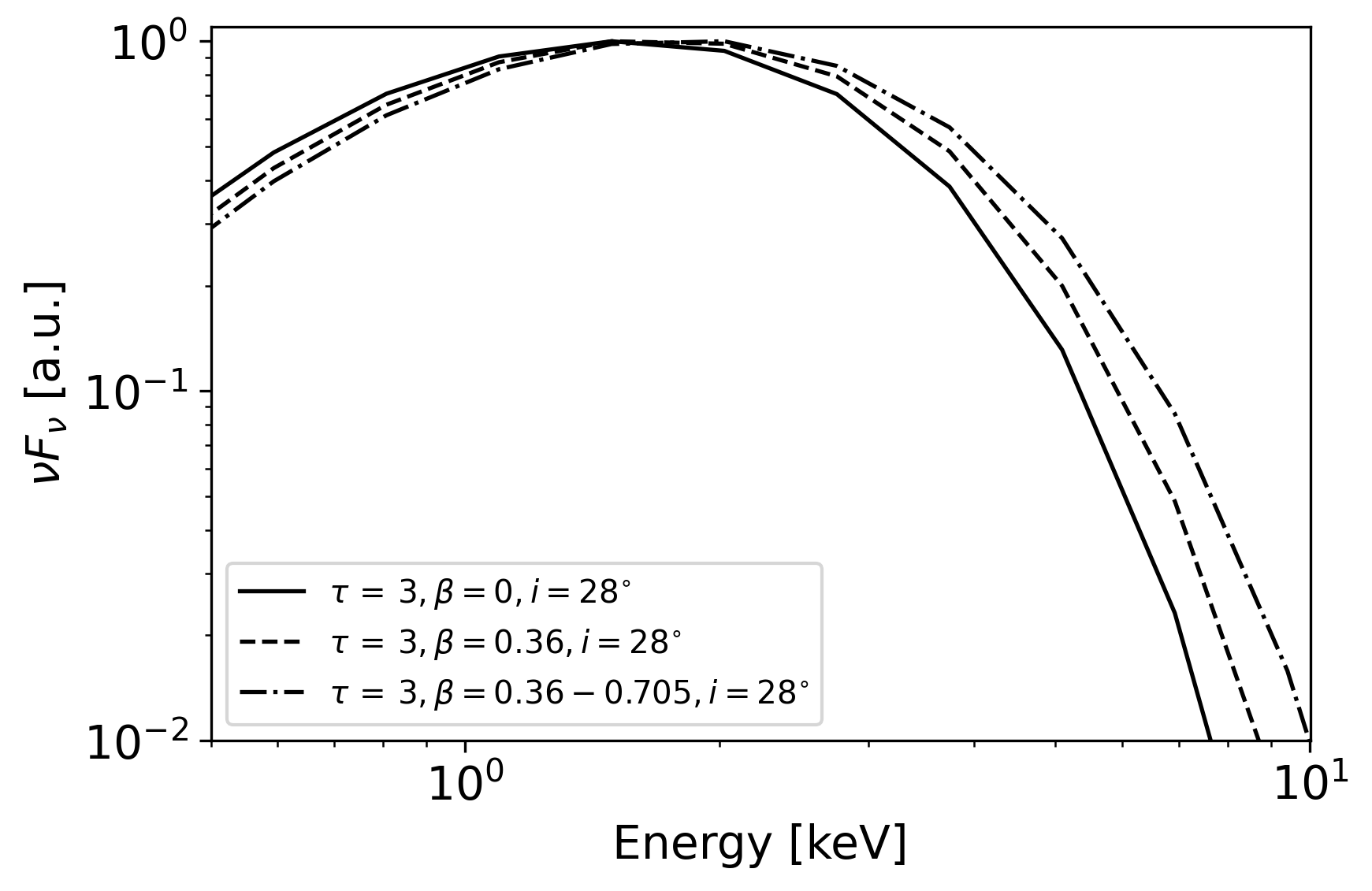}
  \includegraphics[width=0.45\textwidth]{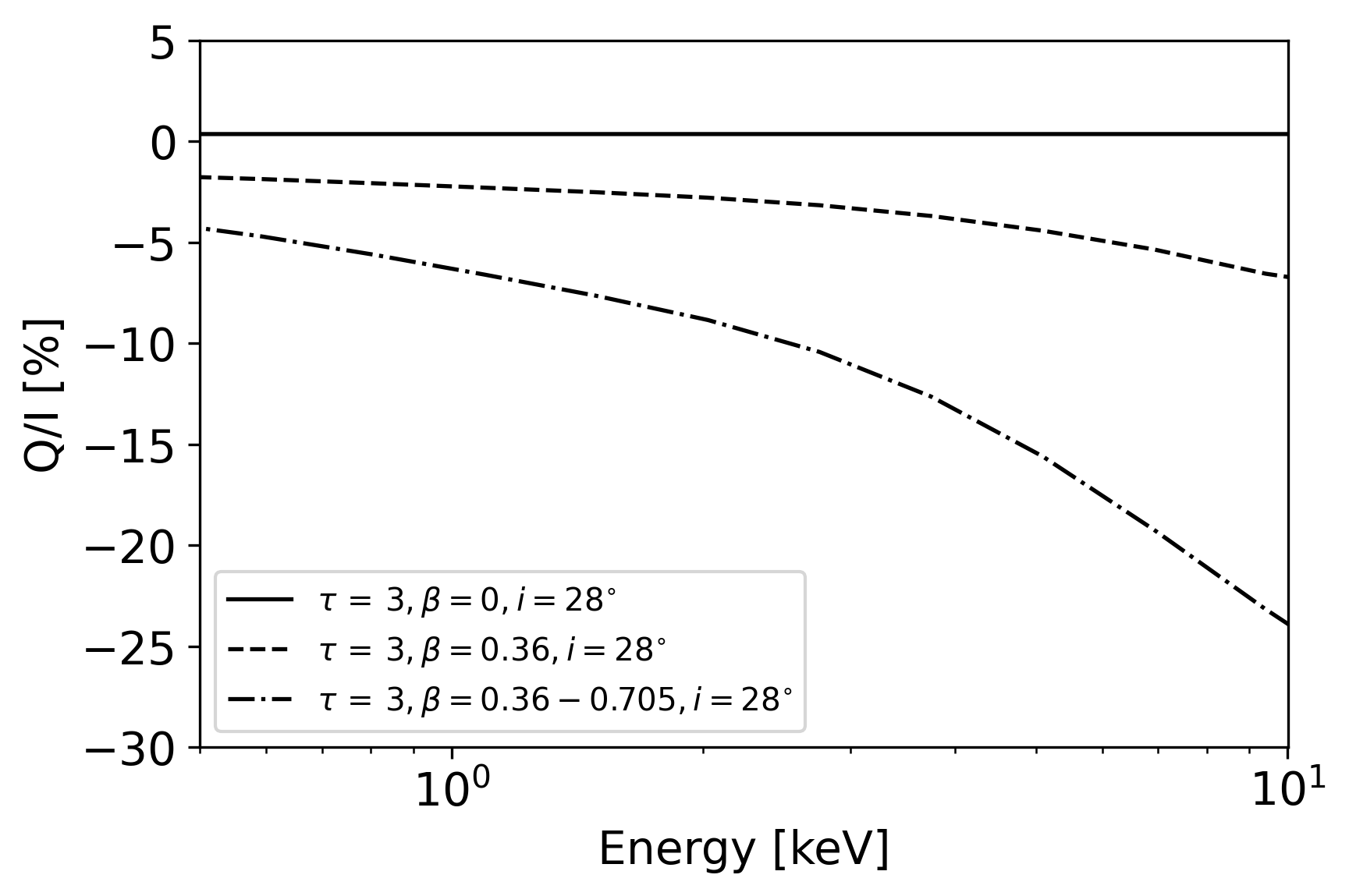}
  \caption{We propose to explain the soft state emission from Cyg X-1 with a new model in which the geometrically thin, optically thick accretion disk is covered with a layer of 
  mildly relativistically moving pair plasma. 
  This figure shows the result from simple radiation transport
  simulations showing the spectral energy distribution (SED, top) and Stokes $Q$ energy spectrum (bottom) for models with different bulk flow velocities $\beta$ in units 
  of speed of light.
  The purple line is for $\beta\,=\,0$, the green line 
  is for constant $\beta\,=\,0.36$, and the 
  red line is for the $\beta$-profile of Fig.\,3 in AB98 with $\beta$ varying from 0.36 to 0.705 along the flow.
  The SEDs have been normalized to 1 at their peak.
  Positive $Q$-values correspond to a PA parallel to the surface of the atmosphere, negative $Q$-values correspond to a PA parallel to the surface normal.}
  \label{f:outflow1}
\end{figure}
\begin{figure}
  \centering
  \includegraphics[width=0.45\textwidth]{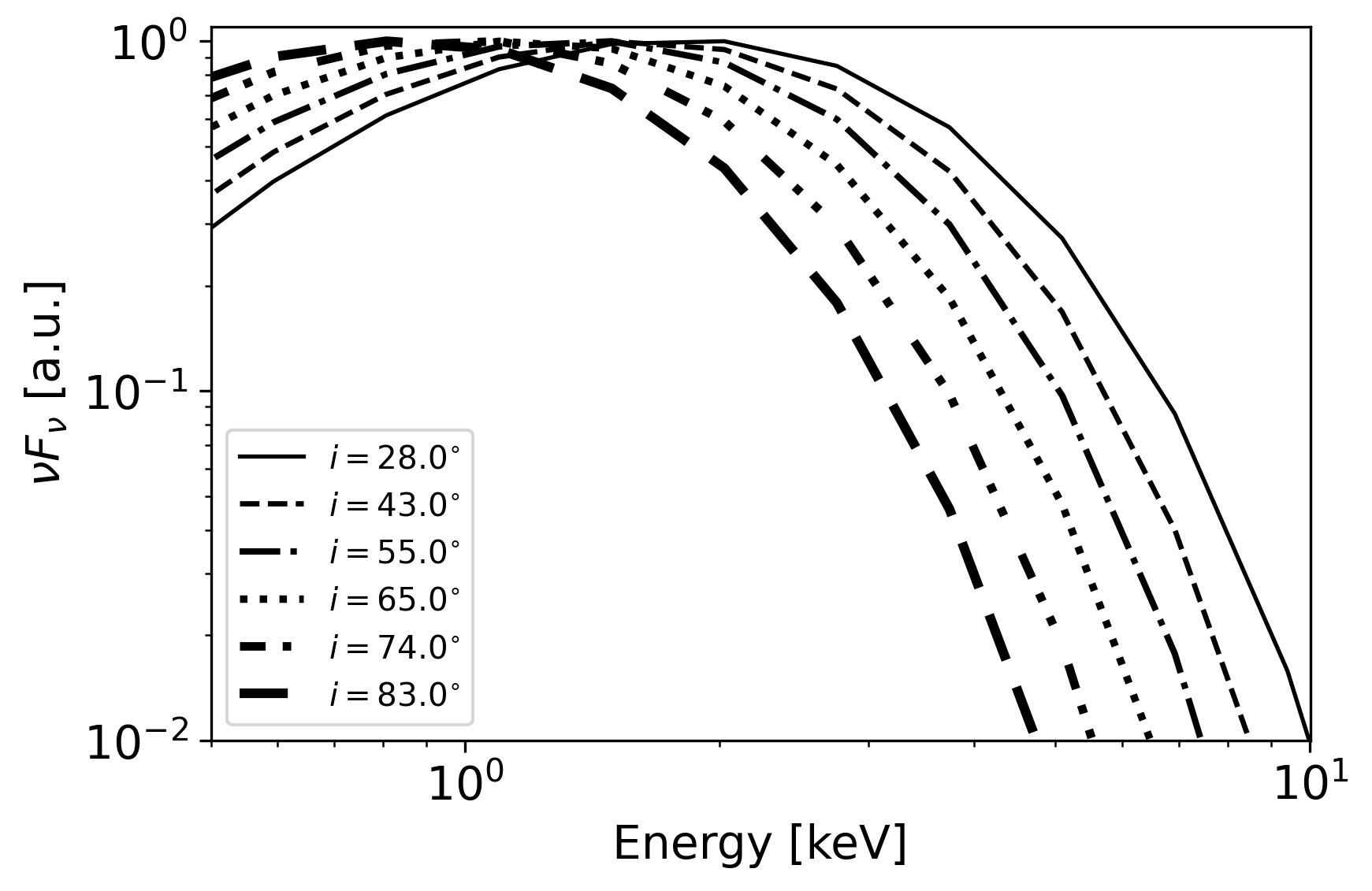}
  \includegraphics[width=0.45\textwidth]{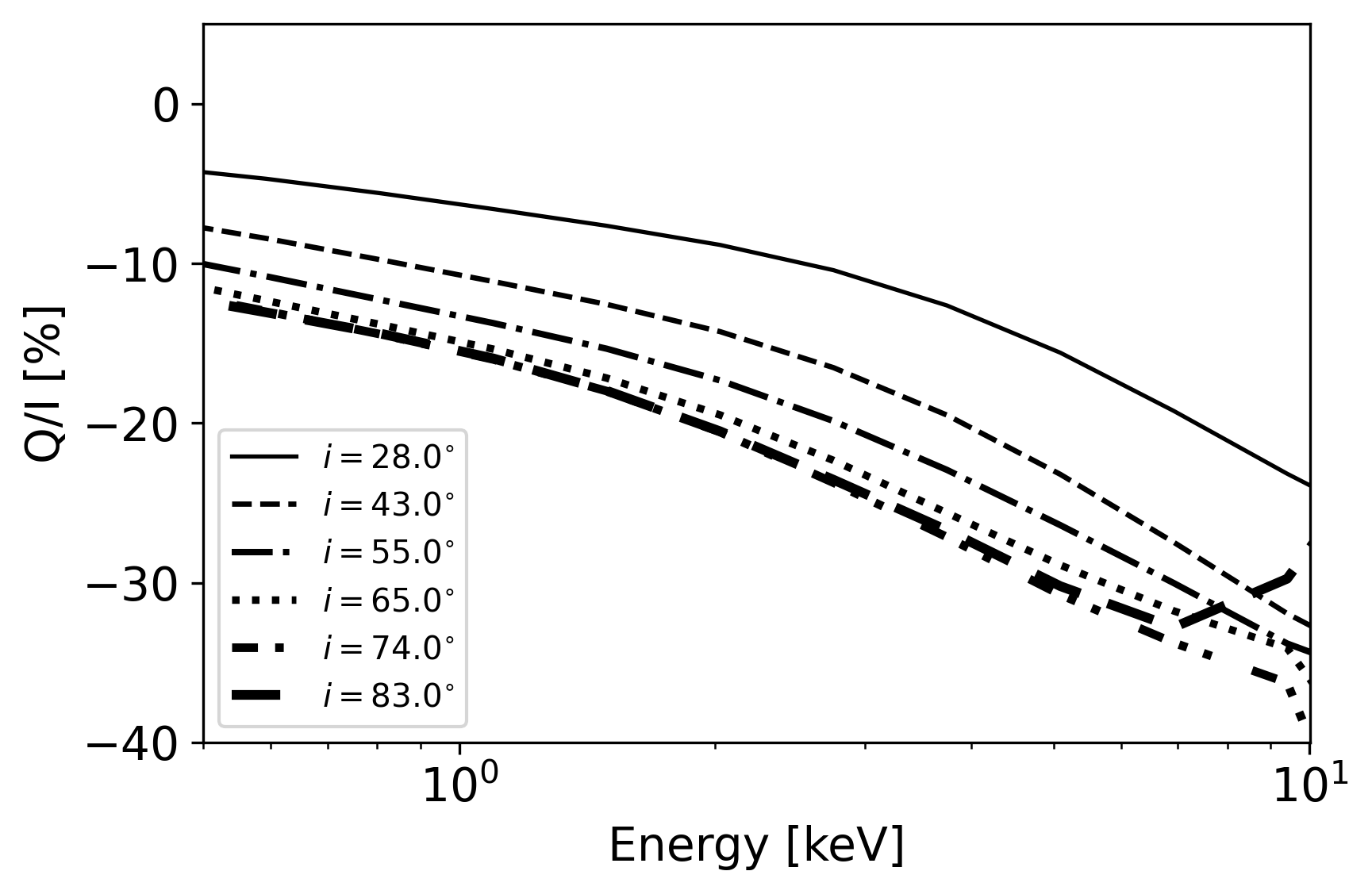}
  \caption{Same as Fig.\,\ref{f:outflow1} but for the model with the $\beta$-profile from Figure 3 of AB98 
  for different inclinations.}
  \label{f:outflow2}
\end{figure}
\begin{figure}
  \centering
  \includegraphics[width=0.45\textwidth]{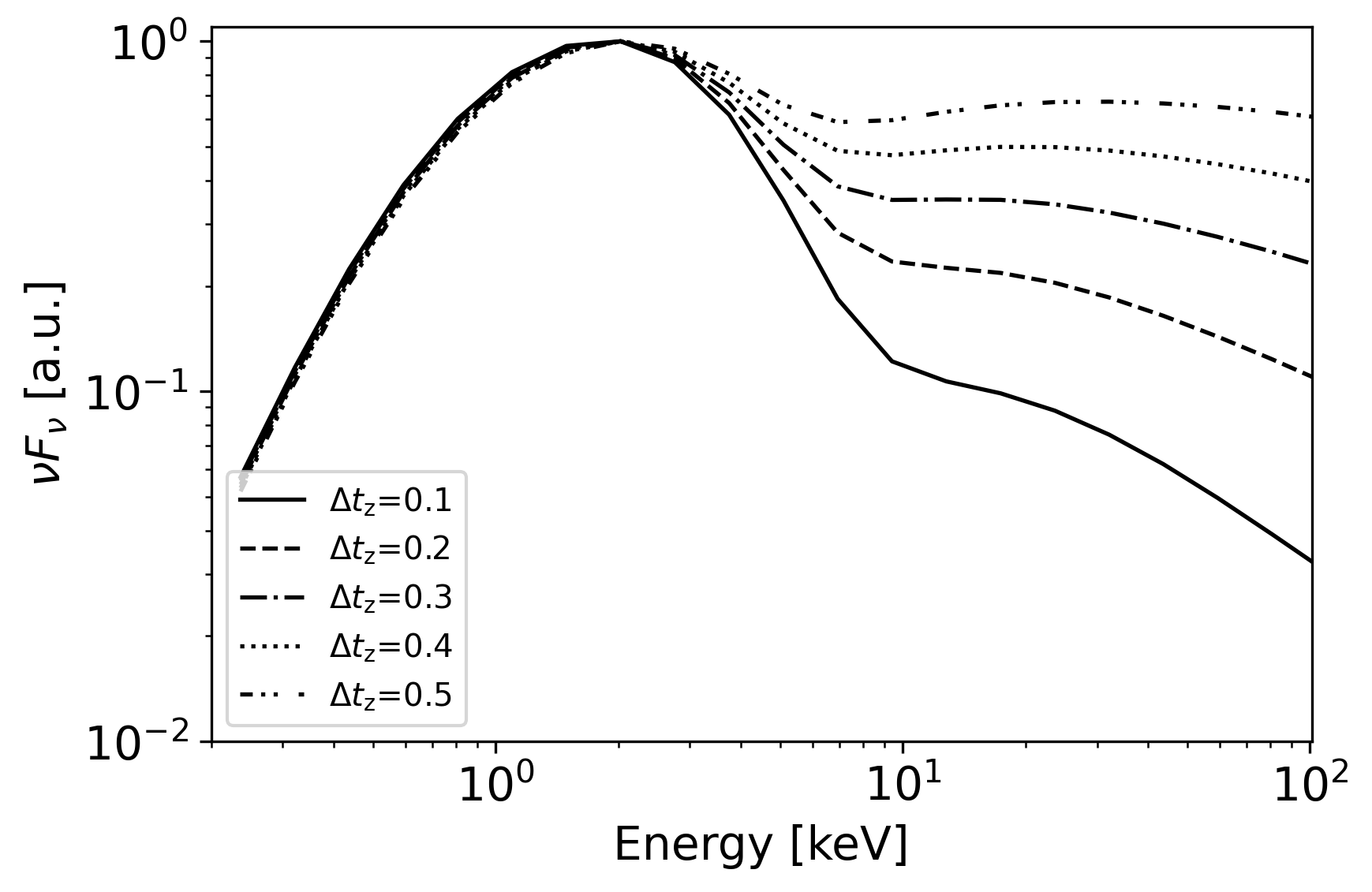}
  \includegraphics[width=0.45\textwidth]{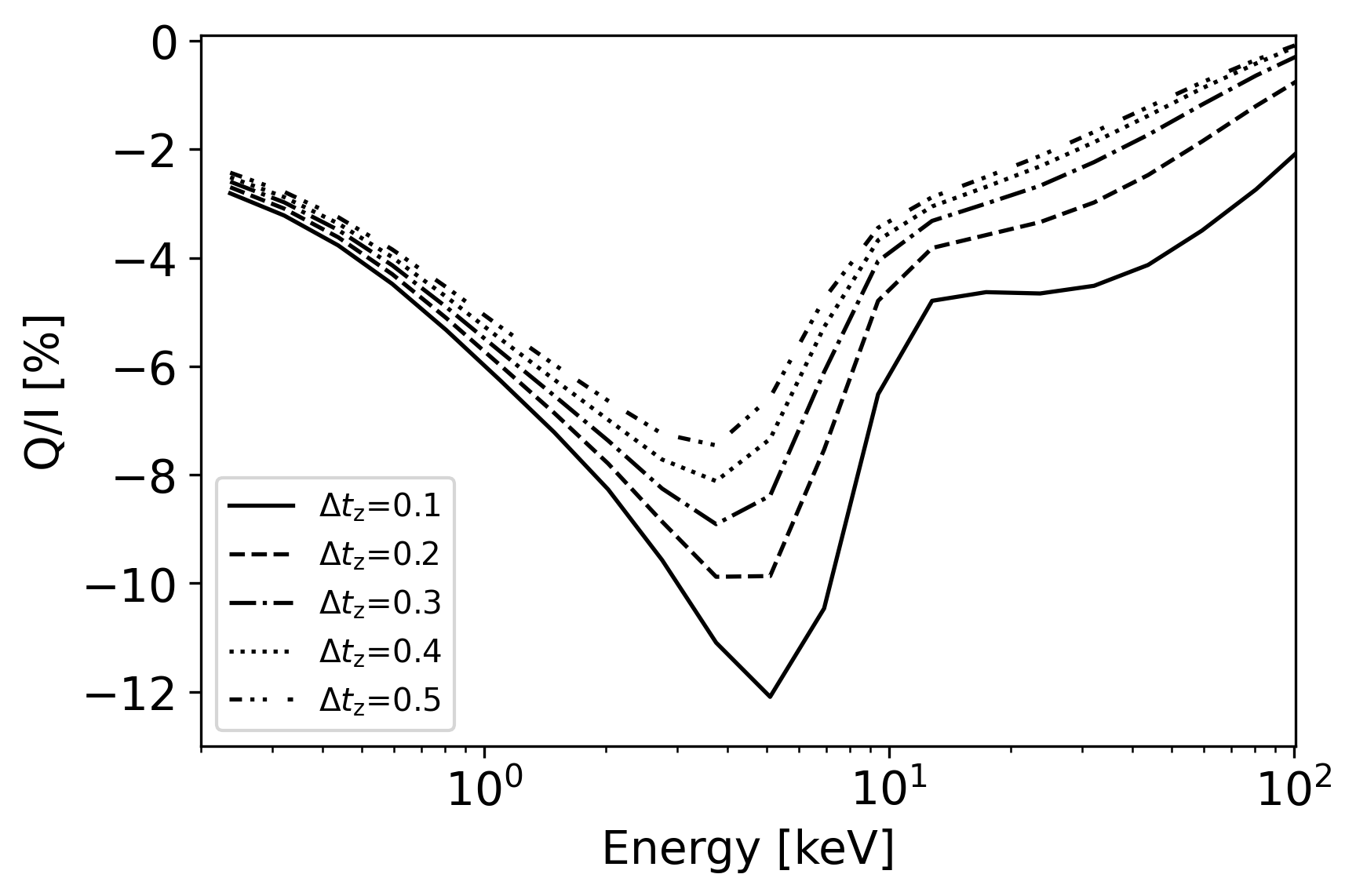}
  \caption{Same as Fig.\,\ref{f:outflow1} with the $\beta$-profile of Figure 3 of AB98 with most of the atmosphere being at 0.3 keV, but with the uppermost layer being at 150 keV to produce an SED similar as during the {\it IXPE} soft state observations of Cyg X-1.
  The different curves are for different 
  uppermost layer optical depths\,$\Delta t_{\rm z}$.
  }
  \label{f:outflow3}
\end{figure}

Figure\,\ref{f:outflow1} presents results for the thermal emission with $T_{\rm i}\,=\,0.3$\,keV  passed through pair plasma of the same temperature $T_{\rm p}$ with various thicknesses $t_{\rm z}$ and $\beta$-profiles for the Cyg X-1 binary inclination of $i\,=\,27^{\circ}.51$.
Note that the model with a $t_{\rm z_{\rm max}}\,=\,3$ and $\beta\,=\,0$ reproduces the PDs from the classical treatment by Chandrasekhar \citep{1960ratr.book.....C}, validating aspects of the code.  
For the rather small inclination of Cyg X-1, the energy spectrum is slightly hardened, but not by much.
As a result, the Comptonizing layer would not distort 
the overall energy spectrum by much, and could still account for the soft state multi-temperature blackbody-type energy spectrum. 
Whereas the spectrum is only slightly modified, the polarization properties change drastically.
The PDs are much higher than in the standard model, and the PA is now parallel rather than perpendicular to the surface normal.
The PDs strongly increase with energy 
as a result of the scatterings required 
to raise the energy of these photons.
The model can create very high PDs parallel 
to the surface normal comparable or even 
exceeding the values observed for Cyg X-1. 

Figure\,\ref{f:outflow2} presents the results 
for the $t_{\rm z_{\rm max}}\,=\,3$ model 
for different inclinations. 
At higher inclinations, the SED peaks at lower energies, and even higher PDs are found.

Figure\,\ref{f:outflow3} shows the results when the uppermost layer of the expanding pair atmosphere is at a higher temperature of $T_{\rm high}\,=\,150$\,keV. 
The layer of coronal plasma takes the PDs down closer to the observed levels. The model with $\Delta t_{\rm z}\,=\,0.2$ produces an SED similar to the one measured during the 
{\it IXPE} soft state observations of Cyg X-1, but the
PDs are still a bit too high compared to the {\it IXPE} 
results, indicating that the plasma is 
moving slower than assumed here. 
The model predicts a marked energy dependence 
of the PDs and could be tested with precision 
measurements of the PDs in the broader 
2-20\,keV energy range. 
The comparison of more detailed modeling of the actual
{\it IXPE} data is outside the scope of this paper.

We do not model the hard state here but refer the reader to \citet{2023ApJ...949L..10P}.

Note that we use the $\beta$-profile from AB98 derived for Thomson scatterings even though our code uses the full Klein-Nishina cross section. We do not expect that the full cross section would change the $\beta$-profile noticeably.

It should be mentioned that actual $\beta$-profiles would depend sensitively on the location of the pair plasma in the accretion flow. Within a few $r_{\rm g}$ of the event horizon, strong gravity and the flux of strongly lensed photons would likely result in a slower upward motion of the pairs. 
\section{Summary and Discussion}
\label{disc}
In the sections above, we emphasized a few important key parameters related to the mass accretion of the Cyg X-1 black hole, i.e., the mass capture rate $\dot{M}_{\rm capt}$, the distance range $r_1-r_2$ over which the soft state and hard state emission draw their energy, the efficiencies $\eta_{\rm grav}$ 
and
$\eta_{\rm BZ}$ 
of converting gravitational 
energy and the spin energy of the black hole into free energy, respectively, and the efficiency of converting 
free energy into X-rays 
$\eta_{\rm rad}$.
A compact corona feeding on the energy of the material within $\sim 5\,r_{\rm g}$ from the black hole can make use of a larger 
fraction of the gravitational energy of the accreting material
$\dot{M}\,c^2$ and can 
tap into the rotational 
energy of the black hole via the BZ mechanism, and is thus energetically favored over an extended corona feeding on the energy liberated at distances between $10\,r_{\rm g}$ and $130\,r_{\rm g}$ from the black hole. 
The former corona has a large scattering compactness and Compton cooling times well below the light crossing times, which
indicates the need for energy transport via Poynting flux or bulk plasma motion.
In contrast, the latter corona has 
Compton cooling times exceeding the 
light crossing times by a few, so 
that sufficiently fast hot plasma 
from the accretion 
disk could power the corona.

We emphasized that the high PDs of Cyg X-1 parallel to the radio jet argue in favor of the existence of outflowing pair plasma in both the soft and hard states.  
A layer of relativistically moving pairs could also explain the high 4-6\% PDs of 4U\,1630$-$47 in the soft state \citep{2024ApJ...964...77R,2024ApJ...977L..10K}. 

The fact that this pair plasma is present in the soft state is highly model-constraining. The Comptonizing plasma needs 
to cover most of the disk. Comptonization in distant  parts of the outflow, as proposed to explain  the hard state emission \citep{2024Ap&SS.369...68M,2024MNRAS.528L.157D}, 
is unlikely to work for explaining the soft state emission, as it would require excessive fine tuning to reproduce the multi-temperature blackbody emission. Furthermore, the Comptonized emission would not outshine the direct emission from the accretion disk.

We propose that shear stresses at the surface of the disk from the disk torquing the plasma above it, the differential rotation of the disk, or the BZ effect power the reconnection that creates the pair plasma. 
The resistive general relativistic magnetohydrodynamic (rGRMHD) simulations of \citet{2025ApJ...979..199S} indicate that the transition 
region between the disk and more tenuous material 
above the disk is a prime location for driving 
dissipation via magnetic reconnection and turbulence. 
The pair plasma would not be easily detectable 
via the 511\,keV emission line, 
as thermal as well as relativistic gravitational and Doppler broadening would give the line a large width. 

If such pair plasmas indeed exist, they would require revising many of the previously derived results, including black hole inclination and spin constraints derived from the continuum fitting method \citep[e.g.,][]{2014ApJ...790...29G}, see also the results in 
\citep{2024ApJ...962..101Z,2024ApJ...967L...9Z},
from Fe K-$\alpha$ line profile studies,  
and from X-ray spectro-polarimetry. 

As mentioned above, the hard state X-ray polarization observations likely indicate an outflowing, laterally extended, possibly structured (patchy) corona inside a 
truncated disk, sandwiching a disk, 
or sandwiching a clumpy flow.
Numerical studies give some first indications 
of possible flow configurations
\citep{2021ApJ...919L..20D,2022ApJ...935L...1L,2024ApJ...966...47L,2025arXiv250508855N,2025ApJ...982..128L,2025ApJ...979..199S}.  
The plasma physics responsible for dissipating the energy is expected to have a major impact on the polarization signatures. 
Plasmoids ejected in the planes of the current sheets  
\citep[AB17,][]{2020ApJ...899...52S,
2021MNRAS.507.5625S,
2023MNRAS.518.1301S,
2024MNRAS.527.6065G}
would impact the polarization signal strongly owing to the strong anisotropy of their bulk velocities. Similarly, turbulent dissipation \citep{2024PhRvL.132h5202G,2024NatCo..15.7026N} can create anisotropies of the velocity vectors of the Comptonizing plasmoids and particles. 
The beamed anisotropic emission can irradiate
certain portions of the accretion disk 
or the observer and would likely be strongly polarized.

The flux and polarization energy spectra,
the reflection ratio (the ratio of the direct and reflected coronal emission), 
the strength and shape of the relativistically broadened Fe K-$\alpha$ line, and the time lags between the fluxes at different energies  depend on the shape and location of the corona
\citep[e.g.,][]{2017MNRAS.472.1932G}, its bulk velocity (impacting the beaming pattern of Comptonized emission toward the disk and toward the observer)  
\citep{1998MNRAS.299..449W,1999ApJ...510L.123B,2015MNRAS.454.4440W}, 
the geometry and ionization state of the 
reflecting plasma \citep[e.g.,][]{2024cosp...45.1519N}, 
and the fraction of disk and coronal emission returning to the disk owing to strong gravitational lensing
\citep{2010ApJ...712..908S,Riaz_2021,2022MNRAS.514.3965D,2022ApJ...934....4K,2025PhRvD.111f3025H}.
The modeling of such data can thus, in principle, constrain these parameters.
For example, the relativistic motion of the corona leads to a reduction of the reflection fraction \citep{1999ApJ...510L.123B}.
In practice, such studies are cumbersome because of the high dimensionality of the parameter space. 
A comprehensive analysis of these effects would be necessary to estimate systematic errors on black hole spin estimates, such as those from \citet[e.g.,][]{2024ApJ...969...40D}, and on 
tests of General Relativity \citep[e.g.,][and references therein]{2018GReGr..50..100K,2024PPN....55.1420B}.

For approximately 60 yr, astrophysicists have been working on constraining the properties of accretion flows onto mass accreting black holes. Unfortunately, the system is still observationally under-constrained. 
We thus caution against claims that we have already identified the correct model. 
It will be  important to keep all viable models in play until additional observations 
and  higher-fidelity numerical modeling 
will constrain the properties 
of the accretion flow further.
\begin{contribution}
H.K.\ wrote most of the text of this paper as well as the code for the the figures in this paper. 
K.H.\ tested and improved the Comptonization engine of the code and made comparisons of the results of the code with published results. 
K.H.\ clarified the definition of the
optical depth used in the paper AB98 
and the special relativistic 
transformations of the various quantities
entering their optical depth. 
K.H.\ furthermore contributed with 
comments throughout the paper.
\end{contribution}
\begin{acknowledgments}
The authors thank Andrei Beloborodov for steering them toward his 1998 paper. They thank Daniel Gro{\v{s}}elj, Nicole Rodriguez Cavero, Ephraim Gau, Sohee Chun, Hamta Farrokhi Larijani, Argen Detoito, Shravan Vengalil Menon, Maitreya Kundu, John  Groger, Kristin Liu, Fang Zhou, Matt
Fritts, and Megan Dickson for helpful discussions. The authors thank
Ephraim Gau and Shravan Vengalil Menon
for carefully reading the manuscript 
and very valuable comments.
The authors thank an anonymous referee for very helpful improvement suggestions. 
Discussions with the {\it IXPE} and {\it XL-Calibur} teams are very much acknowledged. 
The authors thank NASA for support under the grants  80NSSC24K1178, 80NSSC24K1749, and 80NSSC24K1819, and acknowledge support from the McDonnell Center for the Space Sciences at Washington University in St. Louis.
The authors acknowledge the use of the High Performance Computing cluster of the WashU Physics Department 
maintained by S.\,Iyer.
\end{acknowledgments}
\bibliography{manuscript}{}
\bibliographystyle{aasjournalv7}

\end{document}